# Quantum-Kit: Simulating Shor's Factorization of 24-Bit Number on Desktop

Archana Tankasala, Hesameddin Ilatikhameneh

*Abstract* - **Quantum-Kit is a graphical desktop application for quantum circuit simulations. Its powerful, memory-efficient computational engine enables large-scale simulations on a desktop. The ability to design hybrid circuits, with both quantum and classical bits and controls, is employed to demonstrate Kitaev's approach to Shor's factorization algorithm. For the first time, Shor's factorization of a 24-bit integer is simulated with Quantum-Kit in a mere 26 minutes on a modest desktop with Intel Core i5 7400T, 2.4GHz and 12GB RAM. While the largest number factorized so far has been a 20-bit integer, requiring 60 qubits and a supercomputer, the hybrid circuit functionality allows the same number to be factorized using Kitaev's trick with only 21 qubits, in 2.3 minutes, on a desktop. Furthermore, conventional Shor's algorithm for a 13-bit integer with 39 qubits is shown to be 35x faster with Quantum-Kit.**

Quantum computing and communication are promising revolutionary technologies that exploit quantum phenomena like superposition and entanglement to achieve an exponential advantage over their classical counterparts [3-8]. Quantum computing has unlocked new applications ranging from logistics and scientific research to encryption and machine learning, currently driven dominantly by digital technologies [9-12]. It shows the potential to break secure encrypted codes and mine big-data exponentially faster than classical computers [13-14]. Several big players in the industry are already heavily invested in the race to make the world's first practical and viable quantum computer. While the quantum hardware development advances, the capability to simulate quantum algorithms is critical in understanding the performance and potential of real quantum machines when they are ready. To successfully understand and build quantum algorithms that solve highly demanding problems, it helps to run experiments on a computer, using numerical modeling and simulations, ideally emulating a real quantum machine.

Quantum-Kit (Q-Kit) is a graphical quantum circuit emulator that is useful in effortlessly constructing complex quantum circuits, analyzing them and visualizing the quantum states after every quantum operation. While there are several existing software freely available for these quantum simulations, they are either focused on learning or on large-scale simulation capabilities. The former kind are equipped with a user-friendly interface - specifically set-up for learning on fewer qubits, while the latter require a knowledge of quantum programming languages to truly exploit the potential of these software on supercomputing clusters.

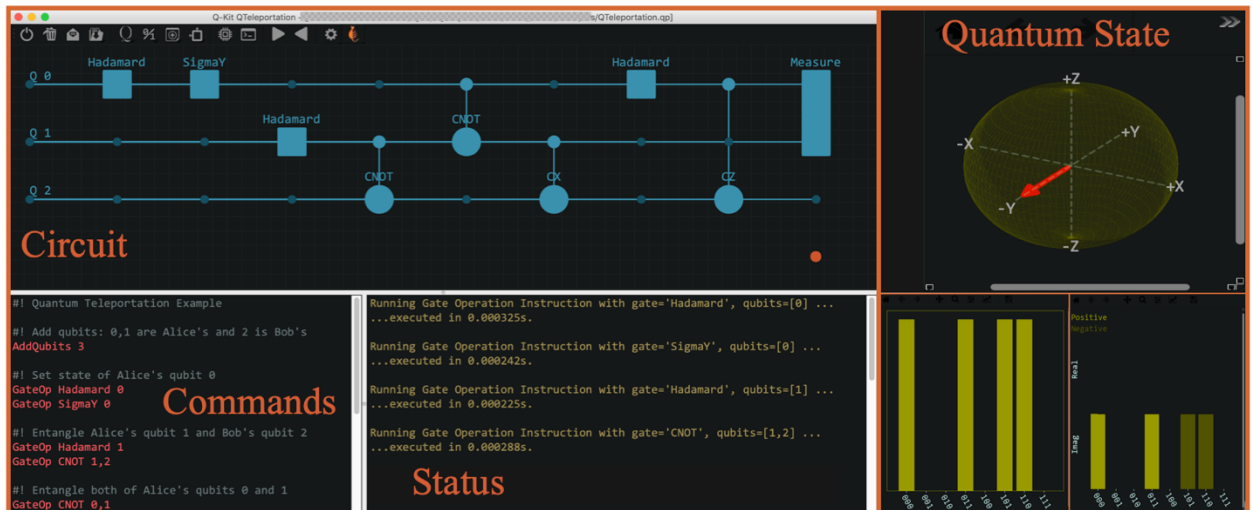

Fig. 1. (a) Quantum-Kit Graphical User-Interface. Panels for (1) graphical quantum **circuit** builder, (2) quantum circuit **command** script, and (3) simulation execution **status**. (b) Visualization of **quantum state** after each quantum gate operation in the circuit. Single qubit states shown on a Bloch Sphere and multi-qubit gates on bar charts for probability distributions and amplitudes.

*Q-Kit.*

Q-Kit is aimed at achieving a balance of the two, a user-friendly interface with capabilities to efficiently run large simulations. Quantum circuits can be set-up in Q-Kit in two different ways. First, using graphical circuit builder, as shown in the top panel of the Q-Kit GUI in Fig. 1(a). Circuits are constructed simply by adding qubits from the menu, assigning quantum gates (single-qubit gates, entangling gates, multi-qubit gates or oracle) by clicking on the qubits, followed by (in-place or deferred) measurement of the qubits.

A second way, especially useful for large quantum circuits, is to set-up with *only* three commands to (1) add qubits, (2) perform a gate operation, and (3) measure, as shown in the bottom left editor panel in Fig. 1(a). An additional command to (4) define custom quantum gates gives users the flexibility to define a new (oracle or unitary) gate as a matrix or as a function. A quantum gate can be defined once and used across circuits. This powerful feature is extremely valuable in employing Q-Kit for research, as this eliminates any dependence on addition of necessary gates to the list of pre-defined gates in Q-Kit. Furthermore, a click-button conversion of the circuit from circuit-diagram to command-script and vice-versa, makes it easy to learn Q-Kit and to edit complex circuits quickly. Commands for very large circuits or quantum programs can be easily generated using simple python scripts as shown in Appendix A.

Status of the running simulations is displayed in the bottom right panel of Q-Kit, shown in Fig.1(a). A thorough logging information on the execution details, including timing analysis, is printed to this panel and saved to a file as well.

At the end of execution, not only is the final quantum state printed to this panel but the quantum state (after every gate operation in the quantum circuit) is also available in the circuit-diagram, as shown in Fig. 1(b). The quantum state can be viewed as (1) probability amplitudes with the complex coefficients, (2) normalized probability distribution, or (3) plain text for user post-processing.

In addition, Q-Kit has several features including the ability to copy data from measured qubits to classical bits and using classical controls for quantum operations. This ability to use classical bits and controls in a quantum circuit, especially in certain algorithms, offers exponential advantages in simulations, both in terms of speed and memory requirements, as discussed in the next section. Q-Kit can be run both in performance-enhanced or memory-enhanced mode and is demonstrated to be extremely efficient although it currently runs only on a single core. A parallelized Q-Kit is expected to enable much more powerful computations, with far fewer resources.

Quantum emulators like Q-Kit are an invaluable tool in an intuitive understanding of the quantum circuits, like quantum teleportation, Shor's factorization or Grover's search algorithms. It allows experimenting different quantum algorithms by modeling qubits on a desktop or laptop computer to study the complex quantum systems on a computer instead of a laboratory. It also allows testing the quantum software on the emulator before running on quantum hardware. Q-Kit has applications not only to research, development and validation of quantum algorithms but also towards an intuitive understanding of quantum circuits, as a high-level precursor to quantum programming.

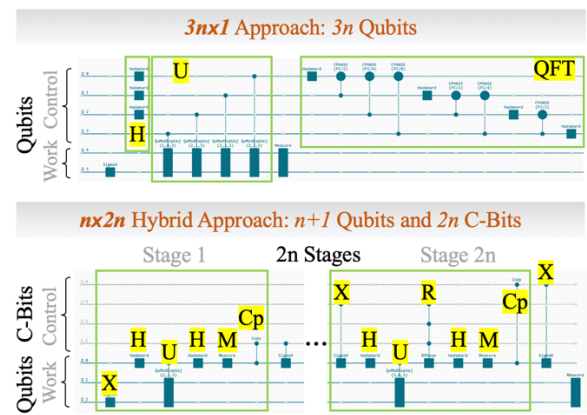

Fig. 2: Schematic representation of Shor's factorization circuit using (a) *3n* qubits in *3nx1* approach and (b) *n+1* qubits and *2n* classical bits in *nx2n* approach. It must be noted that these circuits is only for illustration and the exact circuit for N=15 is presented in Fig. 6. In the *nx2n* approach, the circuit has *2n* stages and the output of the *recycled* qubit at every stage is copied to a classical bit - Cp, and used to initialize the next stage with Sigma X gate. Q-Kit features hybrid circuits, using both classical and quantum bits, which makes this feasible.

*Shor's Factorization with Q-Kit.*

Exponential-time factorization of large semi-prime integers is fundamental to encrypting schemes, like RSA, in securing digital information. Shor's factorization is a polynomial time algorithm that challenges this encryption and has been of great interest, especially with the recent developments in realization of quantum hardware [15-20].

The capabilities of quantum computing simulation software are also most commonly evaluated on Shor's factorization algorithm[1-2,21-22]. While the algorithm is ingenious and offers exponential advantage to a quantum computer against its classical counterpart, classical simulation of the algorithm is extremely computationally demanding. Factorization of a 10-bit number, for example, requires 10 qubits in the work register and 20 qubits in the control register, a total of 30 qubits demanding 32MB of memory. This memory requirement scales exponentially with number of qubits, demanding as much as 32GB for a 15-bit number, clearly not feasible on conventional desktop machines.

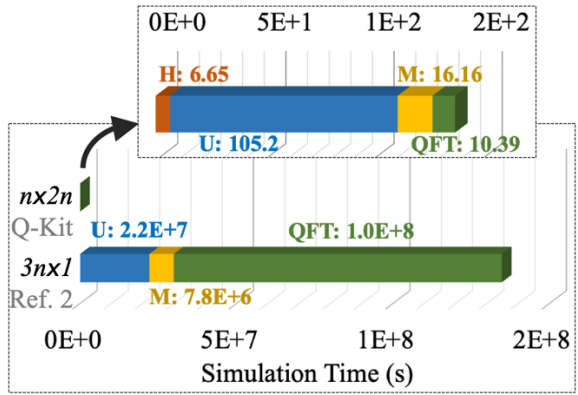

Fig. 3: Timing analysis. Shor's factorization of 20-bit N=961307, with both *nx2n* and *3nx1* approaches, plotted on a LOG scale. As expected, *nx2n* approach is (6) orders of magnitude faster than *3nx1* approach. Data for *nx2n* is from Q-Kit, and for *3nx1* is taken from Ref.[2].

Alternate proposals to Shor's factorization algorithm include Kitaev's version with a clever one-control-qubit trick [23-25], with 10 qubits in the work register and only a single qubit in the control register, which is "recycled" several times. This approach uses a memory of 32KB (instead of 32GB) for factorizing a 15-bit number. However, although the approach is memory efficient and requires fewer qubits, QFT is applied over several stages in the circuit (instead of just once). For the rest of this paper, the former approach with *3n* qubits and a single QFT will be referred to as the *3nx1* approach and the latter approach with (*n*+1) qubits and QFT applied over *2n* steps as the *nx2n* approach.

A schematic illustration of the circuits for *3nx1* and *nx2n* approaches is shown in Fig. 2. In the *3nx1* approach, Hadamard and QFT gates are applied to *2n* control qubits and this makes the quantum state of *3n* qubits a dense vector, demanding huge memory. In contrast, the *nx2n* approach keeps the *n+1* qubit quantum state relatively sparse. However, the latter has *2n* stages and the output of the "recycled" qubit is copied to a classical bit at the end of every stage and is used to initialize the next stage in the circuit. The hybrid circuit feature, with both classical and quantum bits, in Q-Kit allows an easy implementation of the *nx2n* approach, thus making the simulations of Shor's factorization for large numbers feasible.

Several earlier works have demonstrated simulations of Shor's factorization, taking advantage of the specific properties of the algorithm, towards enabling factorization of a 15-bit number (using 45 qubits) on a 16GB desktop. To the best of our knowledge, the largest number factorized so far using quantum simulations is 961307, a 20-bit number that required 37,444 CPU hours and 13.824 TB of memory on a supercomputer.

This work demonstrates simulation of Shor's factorization of the same N=961307 on a 12GB Intel Core i5 desktop in a mere 139s on a single core, using the *nx2n* approach with Q-Kit. Fig. 3 presents a detailed timing analysis for this simulation. It must be noted that this ~$10^6$x speed-up is achieved with Kitaev's *nx2n* approach, unlike Ref.[2]. However, even with exactly the same approach as Ref.[2], Q-Kit is still 35x faster, as shown later for a 13-bit N=8189, a 39-qubit simulation.

Furthermore, for the first time, Q-Kit was employed in simulating Shor's factorization of numbers larger than 20-bits, on a desktop. Up to 24-bit integers are successfully factorized on a 12GB Intel Core i5 machine in ~1555s (26 minutes). Shown in Fig. 4 is the result from 3 runs of Shor's factorization algorithm for 24-bit N=13564597, using the *nx2n* approach.

The log file at the end of simulation is processed to extract a detailed timing report for different quantum gates, over all the *2n* stages of the circuit. The execution times for each gate are shown in Fig. 4 for each of the three runs. Next, the final quantum state the qubits collapse to, after measurement, is used to obtain the order of the quantum state using the method of continued fractions. As seen in Fig. 4, two of the three runs collapse to a state that can successfully estimate the order and therefore, the prime factors of N. It must be noted that although the last run was not able to estimate the prime factors, it does indeed collapse to a right quantum state (order 564840 is a multiple of 141210).

Simulations of numbers larger than N are limited by the available memory. While Q-Kit scales perfectly in memory, the simulation of the above 24-bit number already demands 8GB of memory, close to the desktop limit. 25-bit numbers need machines with larger memory, like supercomputers or AWS.

Fig. 5 demonstrates the significantly improved performance of Q-Kit in simulating Shor's factorization of *n*-bit integers compared against the published data [1,2]. Q-Kit simulations are orders of magnitude faster, because of the *nx2n* approach, than the previously reported values. Even with the exactly same *3nx1* approach (curve 3a), Q-Kit is ~35x faster.

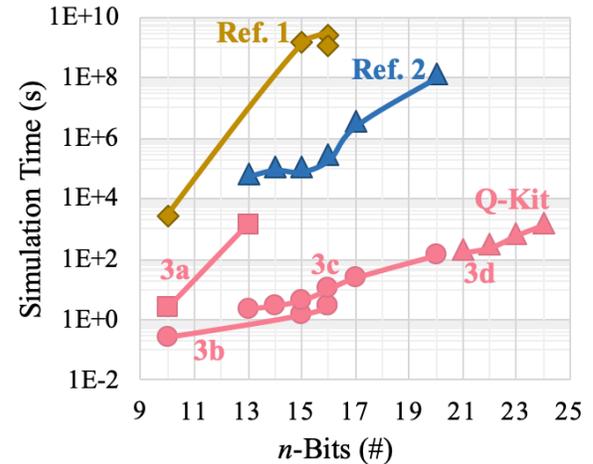

Fig. 5: Performance comparison of Q-Kit in factorizing exactly the same *n*-bit integers as in references [1,2]. Curves 1 and 2 correspond to Ref.[1] and Ref.[2], respectively. Curve 3a is from Q-Kit with the same *3nx1* approach as in Ref. [1,2]. Curves 3b and 3c are also from Q-Kit for exactly the same N as in Ref.[1] and Ref.[2], respectively, but with the *n*x*2n* approach using hybrid circuit functionality. Curve 3d is from Q-Kit for factorization of 21-24 bit numbers, demonstrated here for the first time on a 12GB Intel Core i5 desktop.

*Circuit.*

The quantum circuit corresponding to Shor's factorization algorithm is developed using Q-Kit. Fig. 6a and Fig. 6b show the Q-Kit circuits for factorization of 4-bit N=15, based on *3nx1* and *nx2n*, respectively. In both cases, the measurement is based on random statistical probabilities. Commands for larger circuits can be easily generated from python scripts. Appendix A presents scripts to output commands for a quantum program for both *3nx1* and *nx2n* circuits for any N.

Fig. 4: Q-Kit simulations for Shor's factorization of N=13564597, a 24-bit number on a 12GB Intel Core i5 desktop. Average simulation time is 1555s. Prime factors are extracted from the measured states using continued fractions. 2 of the 3 simulation runs collapse to a measurement that can be used to successfully estimate the order from continued fractions.

Simulations with *3nx1* approach have been described in great detail in several earlier works[1,2]. The circuit uses 2*4=8 qubits in the control register and 4 qubits in the work register, without any compilation optimizations. And the quantum state of the qubits in the control register after measurement is used to extract the order. However, not every collapsed state will give the order with the continued fractions method and a few runs might be required to land on the right state, with either of the approaches, as already shown in Fig. 4.

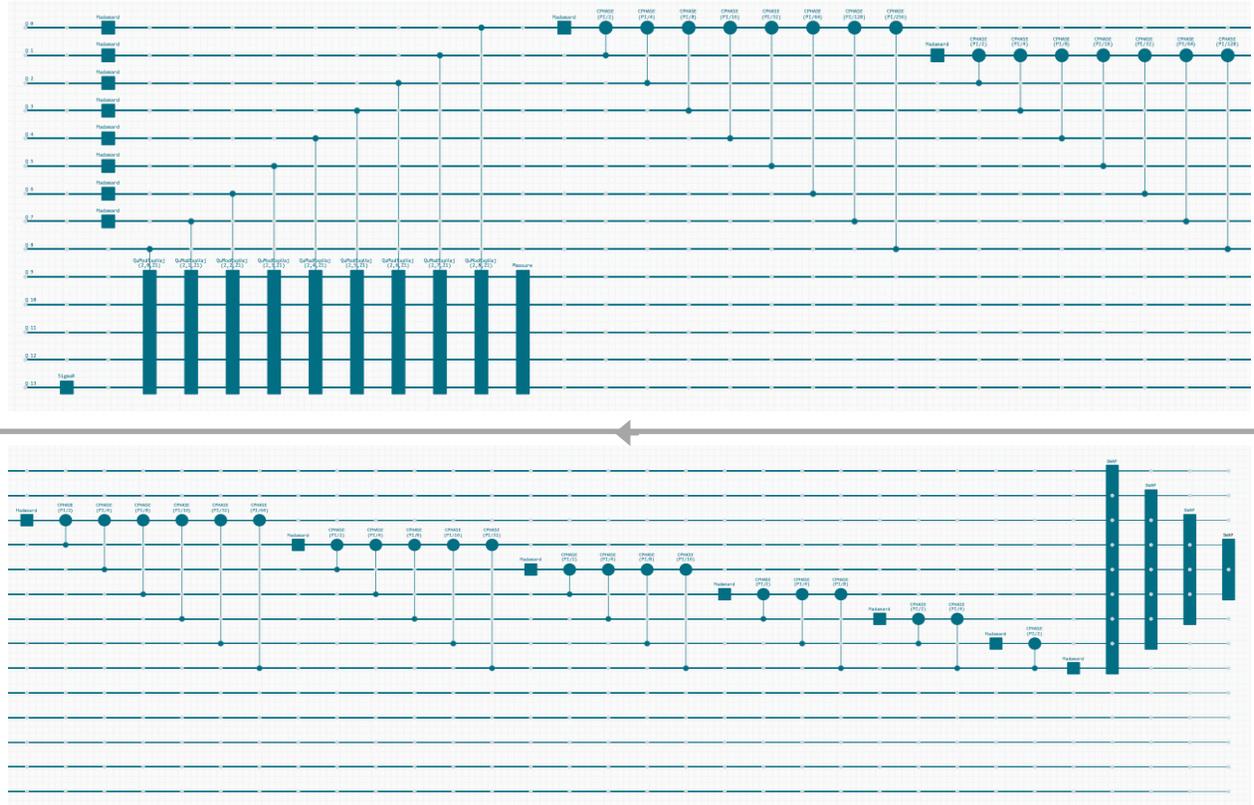

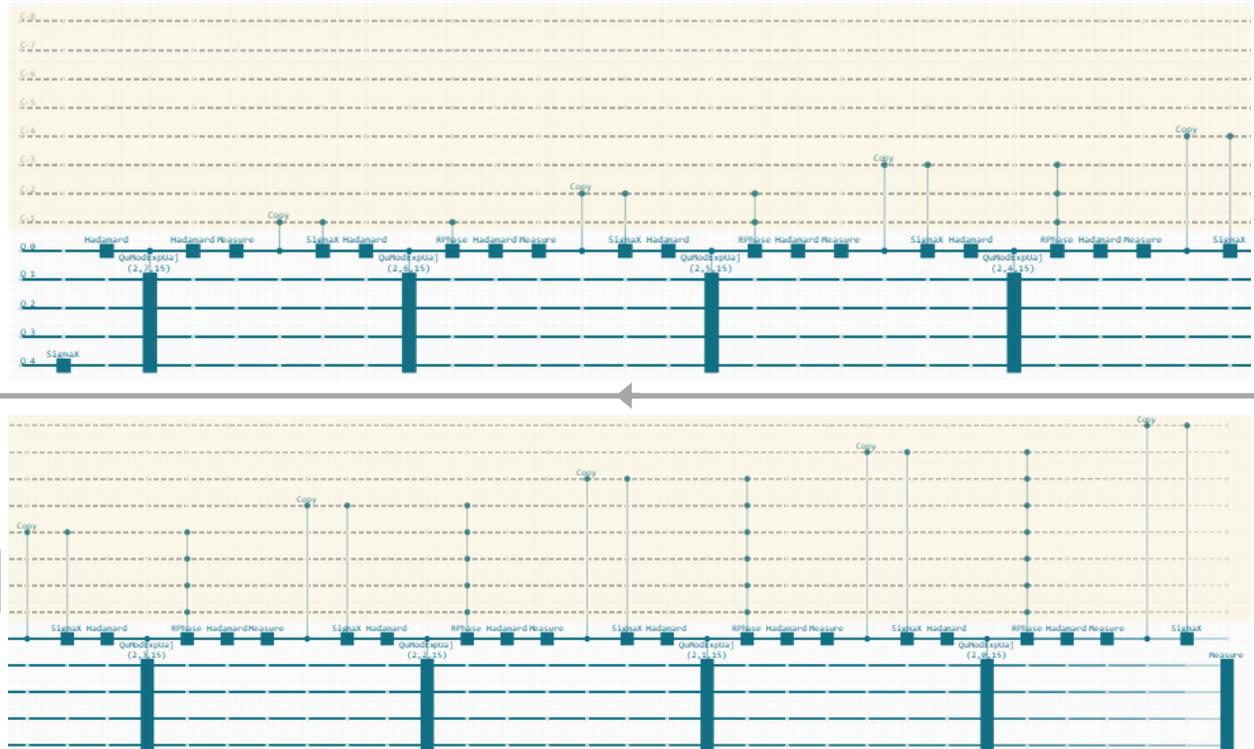

Fig. 6: Quantum circuits for Shor's factorization of 4-bit N=15. (a) Conventional full circuit, *3nx1* approach, with 3*4=12 qubits, 4 in work register and 8 in control register. (b) Kitaev's approach to Shor's factorization, *nx2n* approach, with 4 qubits in work register and a single qubit in control register, with QFT on control register over 8 stages. Each time the measurement of the control qubits is copied to a classical bit (represented by dashed line and highlighted in colored background).

With the simulation based on *nx2n* approach, a single qubit Q0 in Fig. 5b is used repeatedly 2*4=8 times and each time its measured state is saved to one of the 8 classical bits (C-1 to C-8). The classical bits are also used as a control to initialize the qubit for the next stage. The state of the 8 classical bits at the end of simulation corresponds to the final quantum state the system collapses to, as in *3nx1* approach, and this state can be used to extract the order and prime factors using continued fractions.

*Conclusion.*

This work demonstrates the capabilities of Q-Kit, a graphical quantum circuit emulator. It is shown that unlike several existing quantum simulation software, Q-Kit is both user-friendly and computationally powerful. It is equipped with several useful features, including user-defined custom gates and inclusion of classical bits in quantum circuit. This enables simulations of Shor's factorization using Kitaev's one qubit trick, thus requiring fewer qubits, on a desktop as against demanding resources on supercomputers. Furthermore, numbers larger than 20-bit, up to 24-bits, have been successfully simulated in less than 26 minutes for the first-time, on a 12GB Intel Core i5 desktop.

# Appendix A

Example python script to generate quantum program commands for Quantum-Kit simulations of large circuits for Shor's factorization algorithm. The output .qp files from the following script can be simply loaded into Quantum-Kit and executed.

```python
import datetime
from math import log2

def buildShorQP(N=None, a=2, file=None, approach='3nx1'):
    
    #! Output file for Quantum Program.
    if not file:
        file = 'Shor-N'+str(N)+'-a'+str(a)+'-'+approach+'.qp'
    
    #! Timestamp the command generation.
    cmds = []
    cmds.append('#! '+datetime.datetime.now().strftime("%Y-%m-%d %H:%M"))
    
    #! Kitaev's circuit.
    if approach == 'nx2n':
        
        #! Number of quantum and classical bits.
        nQ = N.bit_length()+1
        nC = int(log2(1<<((N**2)-1).bit_length()))
        
        #! Add qubits and classical bits to circuit.
        cmds.append('\n#! Add Qubits and Cbits.')
        cmds.append('AddQubits '+str(nQ))
        cmds.append('AddCbits '+str(nC))
        
        #! Initialization of work register.
        cmds.append('\n#! Initialize work register.')
        cmds.append('GateOp SigmaX '+str(nQ-1))
```

```python
        #! Loop over #stages in Keitev's approach.
        for e in range(nC):

            #! At every stage, apply Hadamard and Quantum Modular Exponentiation.
            cmds.append('\n#! Stage '+str(e))
            cmds.append('GateOp Hadamard 0')
            cmds.append('GateOp QuModExpUaj 0:'+str(nQ)+' a='+str(a)+' j='+str((nC-e-1))+' N='+str(N))

            #! Rotation and Hadamard after QuModExp.
            if e > 0: cmds.append("GateOp RPhase 0,"+','.join([str(i) for i in range(-e,0)]))
            cmds.append('GateOp Hadamard 0')

            #! Measure and copy the qubit data to classical bit.
            cmds.append('Measure 0')
            cmds.append('GateOp Copy 0,-'+str(e+1))

            #! Initialize before next stage.
            cmds.append('GateOp SigmaX 0,-'+str(e+1))

        #! Measurement.
        cmds.append('\n#! Measure all qubits.')
        cmds.append('Measure 1:'+str(nQ))

    #! Full circuit.
    elif approach == '3nx1':

        #! Qubits in work register and control register.
        nWQ = N.bit_length()
        nCQ = int(log2(1<<((N**2)-1).bit_length()))
        nTQ = nWQ+nCQ

        #! Add qubits to circuit.
        cmds.append('\n#! Add Qubits.')
        cmds.append('AddQubits '+str(nTQ))

        #! Initialization of work register.
        cmds.append('\n#! Initialize work register.')
        cmds.append('GateOp SigmaX '+str(nTQ - 1))

        #! Hadamard on control register.
        cmds.append('\n#! Hadamard on control register.')
        cmds.append('GateOp Hadamard 0:'+str(nCQ - 1))

        #! Quantum Modular Exponentiation.
        cmds.append('\n#! Modular Exponentiation of work register.')
        for c in range(nCQ):
            cmds.append('GateOp QuModExpUaj '+str(nCQ-c-1)+','+','.join(
                [str(i) for i in range(nCQ,nTQ)])+' a='+str(a)+' j='+str(c)+' N='+str(N))

        #! Measure work register.
        cmds.append('\n#! Measure work register.')
        cmds.append('Measure '+','.join([str(i) for i in range(nCQ,nTQ)]))

        #! Quantum Fourier Transform.
        cmds.append('\n#! QFT.')
        for c in range(nCQ):
            cmds.append('GateOp Hadamard '+str(c))
            for i, d in enumerate(range(c+1,nCQ)):
                cmds.append('GateOp CPHASE '+str(d)+','+str(c)+' phi=PI/'+str(2**(i+1)))

        #! Swap after QFT.
        cmds.append('\n#! SWAP.')
        for i in range(nCQ//2):
            cmds.append('GateOp SWAP '+str(i)+','+str(nCQ-i-1))

    #! Write commands to .qp file to be loaded directly to Q-Kit.
    filePtr = open(file, 'w')
    filePtr.write('\n'.join(cmds))
    filePtr.close()

#! Run function to build Shor's factorization QP with choice of N and a.
buildShorQP(N=15, a=2, approach='nx2n')
buildShorQP(N=15, a=2, approach='3nx1')
```